\documentstyle[aps,fleqn]{revtex}    
\begin{document}
\title{On the quantum mechanical nature in liquid NMR quantum computing}
\author{G.L.Long$^{1,2,3,4}$, H. Y. Yan$^{1,2}$, Y. S. Li$^{1,2}$, C. C. Tu$^{1,2}$,\\
S. J. Zhu$^1$, D. Ruan$^1$,Y. Sun$^{1,2,5}$, J. X. Tao$^6$,H. M. Chen$^1$ \\       
$^1$Department of Physics, Tsinghua University, Beijing,100084,
P.R. China\\
$^2$ Laboratory for Quantum Computation and Quantum Measurement,\\ Tsinghua University, 
Beijing 100084, P. R. China\\
$^3$ Institute of Theoretical Physics, Chinese Academy of Sciences,\\
Beijing,100080, P.R.China\\
$^4$ Center of Nuclear Theory, National Laboratory of Heavy Ion Physics,\\
Chinese Academy of Sciences, Lanzhou, 730000, P. R. China\\
$^5$ Department of Physics and Astronomy, Tennessee University, Knoxvill, USA\\
$^6$ Department of Chemistry, Tsinghua University, Beijing, 100084, China}
 
\maketitle    
\begin{abstract}
The quantum nature of bulk ensemble NMR quantum computing ---the center of recent heated 
debate, is addressed. Concepts of the mixed state and entanglement are examined, and the 
data in a 2 qubit liquid NMR quantum computation are analyzed. It is pointed out that 
the key problem in the current debate is the understanding of entanglement in a mixed 
state system. The following points are concluded in this Letter: 1)Density matrix 
describes the ``state" of an average particle in an ensemble. It can not describe the 
state of an individual particle in an ensemble in detail; 2) Entanglement is a property 
of the wave function of a quantum particle(such as an molecule in a liquid NMR sample). 
Separability of the density matrix  can not be used to measure the entanglement of mixed 
ensemble; 3)The evolution of states in bulk-ensemble NMR  quantum computation is quantum 
mechanical; 4) The coefficient before the effective pure state density matrix, 
$\epsilon$,  is an measure of the simultaneity of the molecules in an ensemble. It 
reflects the intensity of the NMR signal and has no significance in quantifying the 
entanglement in the bulk ensemble NMR system. We conclude that the liquid NMR quantum 
computation is genuine, not just classical simulations.
\end{abstract}

Computers based on quantum mechanical principle of superposition can do computation much 
faster than  classical computers\cite{r1,r2,r3}.
The proposals\cite{r4,r5} for quantum computing using liquid-state NMR 
have accelerated the experimental studies of quantum computing in demonstrating quantum 
algorithms and quantum error correction 
codes\cite{r6,r7,r8,r9,r9a,r9b,r9c,r9d,r9e,r9f,r9g,r9h,r9i}. 
However this approach has been in the center of a heated 
debate\cite{r10,r11,r12,r13,r14}. The center of the debate is whether liquid NMR quantum 
computations carried out so far are genuine quantum computation. 

To implement a quantum computation, one has to use a quantum mechanical system to 
represent the wave function and to enbody the quantum computation operation. In a liquid 
NMR system at room temperature, the system is in a thermal equilibrium and the molecules 
are not in the same quantum state. One has to construct an effective pure state by 
certain method, exhaustive averaging\cite{r15} for example. After this, the density 
matrix for the system can be written as 
\begin{eqnarray}
\rho=(1-\epsilon)I/d+ \epsilon\rho_1,
\label{e1}
\end{eqnarray}
where $I$ is the identity matrix in $d=2^N$ dimension, and $N$ is the number of qubit. 
$\rho_1$ is the density matrix for a pure state. It is found that\cite{r10,r11,r12,r13} 
when $\epsilon$ in equation (\ref{e1}) is smaller than a critical value which is true 
for all the present NMR experiments carried so far, the density matrix can be decomposed 
into a linear combination of products of the states of the individual qubits. Separable 
density matrix represents separable state which can be written as product of the 
composite particle wave functions and they are unentangled. It was 
argued\cite{r10,r11,r12,r13} that since the density matrix is separable for small 
$\epsilon$, there was no entanglement. The bulk ensemble NMR computation experiments 
carried so far are not genuine quantum computations and can only be considered as 
classical simulations of the quantum computation, a point strongly disagreed by 
Laflamme\cite{r14}. In this Letter we will address this issue by examining the concepts 
of mixed state and the concept of entanglement. We also analyzed the evolution of the 
density matrices in a 2 qubit NMR quantum computation and compare them with prediction 
of quantum mechanics.

Microscopic particle, such as a molecule, obeys the laws of quantum mechanics. Its state 
is fully described by a wave function whose time evolution is the Sch\"{o}dinger 
equation. We do not talk about the state of a particle pure or mixed. Because it is 
always described by a wave function in quantum mechanics\cite{r17}. Pure or mixed state 
refers to an ensemble. By pure state, we mean that all the molecules in an ensemble have 
the same quantum wave function. In other words, every particle in the ensemble can be 
described by the same wave function.  By mixed state, we mean that the molecules in an 
ensemble are not in the same quantum state, but rather has a classical distribution of 
the molecules in different quantum states: $n_1$ particles in state $|\psi_1\rangle$, 
$n_2$ particiles in state $|\psi_2\rangle$ .... Each of the particle in the ensemble 
{\bf is in a quantum state, described by a wave function}.   However, we do 
not have the detailed information for each particle and what we know is just the 
property of an average particle in the ensemble. To describe the  property of an average 
particle in an ensemble, we use the density matrix which was invented by 
Von Neunmann\cite{r16}. In the 2 qubit system such as phosphoric acid  for example, 
there are 4 quantum states, $|00\rangle$, $|01\rangle$,$|10\rangle$ and $|11\rangle$, 
where the first bit represent the nuclear spin of the $^{31}$P and the second bit 
represents the nuclear spin of hydrogen atom. If all the phosphoric acid molecules are 
in the same state, say, $|00\rangle$, then we have a pure state. For liquid NMR sample 
at room temperature, the molecules are in the thermal equilibrium and the ensemble is 
described by a mixed state.

The idea of entanglement was first formulated by Einstein, Podolsky and Rosen\cite{r18}. 
When we say the wave function of a quantum system is entangled we mean that   the wave 
function of the whole system can not be factorized into a product of the wave functions 
of the constituent particles. 
It should be stressed that here the constituent particles are all quantum mechanically 
related one another, such as the nuclear spins of the hydrogen atom and $^{31}$P in a 
phosphoric acid molecule, not the molecules of an ensemble.  

Now we  study the concept of entanglement in a mixed state. First it is easy to see that 
the same density matrix can be prepared in numerous ways. For example, in a two qubit 
system, a density matrix which is 1/4 of a unit matrix can be prepared by putting into 
the ensemble with a quarter of particles in each of the  4 states with spin up and down 
along the $z$-axis:
\begin{eqnarray}
\rho={1\over 4}|00\rangle\langle 00|+{1\over 4}|01\rangle\langle 01|+{1\over 
4}|10\rangle\langle 10|+{1\over 4}|11\rangle\langle 11|.
\label{e2}
\end{eqnarray}
The same density matrix can be obtained by putting a quarter number of particles into 
the ensemble in each of the 4 Bell states. 
\begin{eqnarray}
\rho&=&{1\over 2}(|00\rangle+|11\rangle)(\langle 00|+\langle 11|)
+{1\over 2}(|00\rangle-|11\rangle)(\langle 00|-\langle 11|)\nonumber\\
&&+{1\over 2}(|01\rangle+|10\rangle)(\langle 01|+\langle 10|)
+{1\over 2}(|01\rangle-|10\rangle)(\langle 01|-\langle 10|).
\label{e3}
\end{eqnarray}
Though the density matrices are the same, the physical systems are completely different. 
There are infinite number of ways to prepare a system with a given density matrix. This 
is not surprising, since density matrix describes only the average property of a system. 
In this example, both ensembles can be viewed as an ensemble consisting of 4 
sub-ensembles with each sub-ensemble in a pure
quantum state. The molecules in these ensembles have quite different entanglement 
properties.  In (\ref{e2}), each individual molecule wave function is factorisable, 
whereas in (\ref{e3}), each individual molecule is non-separable and is enatngled. Since 
density matrix for a mixed ensemble describes only the average property and different 
physical systems can have the same density matrix, one can not tell whether a mixed 
state ensemble is entangled or not.
{\bf Even if a density matrix is separable, it does not mean neccesarily the 
molecules in the ensemble is entangled unless the system is prepared in the way as 
suggested by the decomposition}. Two ensembles such as (\ref{e2}) and (\ref{e3}) may 
have identical density matrices, but the physical systems are completely different, 
though we do not have any measurement to distinguish them. From the mathematical 
properties of the density matrix alone, one can not draw conclusions about the detailed 
properties of the individual particles of the system. The physical properties of the 
ensemble depend on the  history or the method of its preparation \cite{r19}. The pure 
state is an exception, because all the molecules are in the same quantum state, the 
average property of the system is identical to the property of an individual 
particle\cite{r19a}.

Secondly, even we know the history of the  ensemble and locate the physical system,  the 
density matrix of an ensemble 
 reflect only the properties of an average particle in an ensemble and  does not tell 
the information of an  individual particle in the ensemble. From the density matrix, we 
can not say more  about a particle  than its average property.   Taking for example the 
systems represented by (\ref{e2}) and (\ref{e3})(Here we use the expansion in the 
equation to specify its preparation procedure.) The difference in the properties of the 
invidual particles in the two ensembles can not be reflected in the density matrix.

Thirdly, entanglement is a concept for a quantum system only, a single molecule for 
instance. We can talk about the entanglement of wave functions of nuclear spins of the 
hydrogen atom and the $^{31}$P atom of a single phosphoric molecule. But we can not talk 
about the entanglement properties between two molecules in an ensemble. This is true for 
a pure ensemble.
What is meant by entanglement is the entanglement within each individual molecule,{\bf 
not the entanglement between different molecules in the ensemble, and the molecules in 
the ensemble are independent of one another}. The purity of the state is a measure of 
purity of the ensemble, or the synchronization or simultaneity  of the wave functions of 
the molecules in the ensemble. When quantum computation is performed in a pure 
ensemble(a pure state), each molecule in the ensemble undergoes the same evolution 
simultaneously. For a mixed ensemble, the quantum computation operation is performed on 
molecules in different quantum states, including the state we are interested in. 
Effective pure state preparation procedure is only a way to extract the required signal. 
In a simple example, let's look at a one qubit system. Suppose the density matrix of the 
system can be written as ${\rho}=\left[\begin{array}{cc}{5\over 8} & 0\\ 0& {3\over 
8}\end{array}\right]$. In a extremely simplified illustration, we can imagine that there 
are 8 2-level particle. Five of them are in state $|0\rangle$ and three particles are in 
state $|1\rangle$. When we measure the system which is a radio frequency pulse in NMR, 
both $|0\rangle\rightarrow|1\rangle$(upward) and 
$|1\rangle\rightarrow|0\rangle$(downward) transitions are induced.  The transition 
signals meet in the coil, 3 upward transitions cancel with 3 downward transitions. The 
net signal in the coil is 2 upward transitions. This is equivalent to a pure system with 
two particles in state $|0\rangle$. When the number of qubits becomes large, effective 
pure state procedures are used to make unwanted signals cancel and the required signals 
collected. From this simple picture we see  $\epsilon$, the coefficient before the 
effective pure state density in (\ref{e1}) is just a measure of the number of molecules 
in the required state whose signal can be read out. It is closely related to the 
intensity of the readout signal. There is no essential difference between an NMR 
experiment with $\epsilon=1$ and one with a smaller $\epsilon$ value.

In a liquid NMR experiment, the evolution of the molecules in the ensemble are quantum 
mechanical. We have analyzed the experimental data between two steps in a phase matching 
study of the generalized quantum searching algorithm\cite{r21}. The details of the 
quantum algorithm is not important here. We give here the transformation matrix for a 
step in the quantum computation is
\begin{eqnarray}
c=\left[\begin{array}{rrrr} 
{3\over 4}+{I\over 4} & -{1\over 4}+{I\over 4}& -{1\over 4}+{I\over 4} & {1\over 
4}+{I\over 4}\\
-{1\over 4}+{I\over 4} & {3\over 4}+{I\over 4}& -{1\over 4}+{I\over 4} & {1\over 
4}+{I\over 4}\\
-{1\over 4}+{I\over 4} & -{1\over 4}+{I\over 4}& {3\over 4}+{I\over 4} & {1\over 
4}+{I\over 4}\\
-{1\over 4}+{I\over 4} & -{1\over 4}+{I\over 4}& -{1\over 4}+{I\over 4} & {1\over 
4}{3I\over 4}\end{array}\right].
\end{eqnarray}
 What we are interested in is the evolution of the density matrix. The density matrix is 
obtained by the quantum state tomography technique\cite{r23} directly from experiment,
\begin{eqnarray}
\rho(1)=\left[\begin{array}{rrrr}
0.1794 & 0.1591 +0.0208 I & 0.0601-0.0001 I & -0.0483-0.0549 I\\
0.1591 -0.0208 I & 0.2453 & 0.1247-0.0281 I & -0.0514-0.1534 I\\
0.0601+0.0001I  & 0.1247+0.0281I & 0.3616 & 0.0099+0.0682I\\
-0.0483+0.0549I & -0.0514+0.1534I & 0.0099-0.0682I & 0.2137\end{array}
\right].
\end{eqnarray}
After one step of quantum computation manipulation, the density matrix is constructed by 
quantum state tomography again,
\begin{eqnarray}
\rho'=\left[\begin{array}{rrrr}
0.2278 & 0.0858+0.0186 I & 0.0640+0.0387 I & 0.0691-0.0372 I\\
0.0858-0.0186I & 0.1006 & 0.1019-0.0062 I & 0.1650-0.0893 I\\
0.0640-0.0387I  & 0.1019+0.0062I & 0.3921 & 0.0454-0.0111I\\
0.0691+0.0372I & 0.1650+0.0893I & 0.0454+0.0111 I & 0.2794\end{array}
\right].
\end{eqnarray}
The theoretical prediction of the density matrix using quantum mechanics through 
$\rho'_{th}=c\cdot \rho \cdot c^{\dagger}$ is
\begin{eqnarray}
\rho_{th}'=\left[\begin{array}{rrrr}
0.1849 & 0.0891 +0.0599 I & 0.0758+0.0225 I & 0.1146-0.0439 I\\
0.0891-0.0599 I & 0.0999 & 0.0650-0.0446 I & 0.1377-0.0861 I\\
0.0758-0.0225I  & 0.0650+0.0446I & 0.3876 & 0.0018-0.0083I\\
0.1146+0.0439I & 0.1377+0.0861I & 0.0018+0.0083 I & 0.3277\end{array}
\right].
\end{eqnarray}
The theoretical prediction of the evolution of the density matrix is in good agreement 
with the experimental measurement.

We summarize the main points of this Letter: 1) The density matrix is not unique in 
locating a physical system. To uniquely specify a physical ensemble, one has to include 
the history of the ensemble. Separability of the  density matrix when $\epsilon$ is 
small alone does not mean the molecules in the ensemble are untangled. The molecules are 
entangled only when the ensemble is prepared in a way suggested by the decomposition. 2) 
Entanglement is a property of an individual particle in an ensemble, not a property of 
the ensemble as a whole.  3) The evolution of the density matrix in a liquid NMR system 
is quantum mechanical. 4) The procedures for preparing an effective pure state is only a 
way to extract the proper signals from the ensemble. The coefficient $\epsilon$ of the 
effective pure state does not determine the entanglement property of an ensemble. It is 
only a measure of the intensity of the signals of the required state. This conclusion 
has an important practical implication for NMR quantum computation, a small $\epsilon$ 
does not change the quantum nature of an NMR quantum computation. One can repeat the 
quantum computation for sufficient number of times to effectively enhance $\epsilon$. 
This will help in scale an NMR quantum computer. One can increase the number of 
repetition to compensate for the decrease of signals for large qubit numbers. 5)
Finally we'd like to stress that power of the quantum computation lie both in the 
coherence of states(superposed states) and the ability to perform quantum mechanical 
unitary operation. Entangled state is a special case of the superposed state. During a 
quantum computation, the quantum register is always in a superposed state, and sometimes 
it is entangled and sometimes it is unentangled.

To conclude, the present NMR quantum computation is genuinly quantum mechanical, much 
more than a simulations of a quantum computation. Liquid NMR technique is still a very 
useful tool in experimental studies of quantum computation. 


\begin{thebibliography}{99}
\bibitem{r1} R. Feynman, Int. J. Theor. Phys., 21, 467 (1992).
\bibitem{r2} P. Shor, In Proc. 35$^{th}$ Annual Symposium on Foundations of Computer 
Science, 124, Los Alamitos, CA, 1994. IEEE Computer Society Press.
\bibitem{r3} L. K. Grover, Phys. Rev. Lett. 79, 325 (1997).
\bibitem{r4} D. G. Cory, A.F. Fahmy and T. F. Havel, Proc. Nat. Acad. Sci. USA 94, 
1634(1997). 
\bibitem{r5} N. A. Gershenfeld and I. L. Chuang, Science 275, 350 (1997).
\bibitem{r6} J. A. Jones and M. Mosca and R. H. Hansen, Nature, 393, 344 (1998).
\bibitem{r7} Liping Fu et al, e-print quant-ph/9905083.
\bibitem{r8} I. L. Chuang, L.M. K. Vandersypen, X. Zhou, D.W. Leung and S. Lloyd, Nature 
393,143 (1998).
\bibitem{r9} D. G. Cory et al, Phys. Rev. Lett. 81, 2152(1998).
\bibitem{r9a} D. G. Cory et al, Physica D 120, 82 (1998)
\bibitem{r9b} I. L. Chuang et al, Proc. R. Soc. Lond. A393,143(1998).
\bibitem{r9c} I. L. Chuang et al, Phys. Rev. Lett. 80, 3408 (1998).
\bibitem{r9d} J. A. Jones and M. Mosca, J. Chem. Phys. 109, 1648 (1998).
\bibitem{r9e} R. Laflamme et al, Phil. Trans. Roy. Soc. London A356, 1743(1998).
\bibitem{r9f} N. Linden, H. Barjat and R. Freeman, Chem. Phys. Lett. 296, 61 (1998).
\bibitem{r9g} M. A. Nielsen, E. Knill and R. Laflamme, Nature 396, 52 (1998).
\bibitem{r9h} G.P. Berman et al, Phys. Rev. B 58, 11 570(1998).
\bibitem{r9i} L. J. Schulman and U. Vazirani, e-print quant-ph/9804060.
\bibitem{r10} K. \.{Z}ycczkowski, P. Horodecki, A. Sanpera and M. Lewenstein, 
quantum-ph/9804024.
\bibitem{r11} G. Vidal and R. Tarrach, quantm-ph/9806094.
\bibitem{r12} S.L. Braunstein, C. M. Caves, R. Jozsa, N. Linden, S. Popescu and R. 
Schack, Phys. Rev. Lett. 83, 1054 (1999).
\bibitem{r13} R. Schack and C. M. Caves, Phys. Rev. A 60, 4354 (1999).
\bibitem{r14} R. Laflamme, http://quickreviews.org/
\bibitem{r17} Caution must be made here. When we say that a single particle is in a 
mixed state we are actually making an averaging process. Some times, we talk about the 
average state of a single particle over a period of time, for example, confer to I. L. 
Chuang, {\it Quantum information  and computation: theory and practice}, Ph.D. thesis, 
Stanford University, 1996, P.28-9. Sometimes we average the wave function over external 
degrees of freedom, and yield a mixed state. But at a given instant, a single particle 
is in a state by a wave function. It is just because of lack of knowledge that we 
represent the state by its average property.
\bibitem{r15} E. Knill, I. Chuang and R. Laflamme, Phys. Rev. A57,3348 (1998).
\bibitem{r16} J. von Neumann, G\"{o}ttinger Nachr. 245 and 273 (1927)
\bibitem{r18} A. Einstein, B. Podolsky and N. Rosen, Phys. Rev., 47, 777 (1935)
\bibitem{r19} U. Fano, Rev. Mod. Phys. 29, 74 (1957)
\bibitem{r19a} There is some minor difference between the property of a pure ensemble and that of a single molecule. For instance, a measurement on state $1/2(|00\rangle+|11\rangle)$ will give either $|00\rangle$ or $|11\rangle$ for a single particle, but will leave an pure ensemble into a mixed ensemble with half of the particles in $|00\rangle$ state and another half of the particles in $|11\rangle$ state. We can not have the same result as a single particle, i.e., a result with either the ensemble in $|00\rangle$ or in $|11\rangle$.
\bibitem{r21} G. L. Long, Y. S. Li, W. L. Zhang and L. Niu, Phys. Lett. A262, 27 (1999). 
The NMR experiment results will be published elsewhere.
\bibitem{r23} I.L. Chuang et al, Proc. R. Soc. London, Ser. A 454, 447 (1998).

\end{thebibliography}
\end{document}